\documentclass[12pt]{article}
\usepackage{graphicx}

\newcommand{\beq}{\begin{equation}}
\newcommand{\eeq}{\end{equation}}
\newcommand{\beqa}{\begin{eqnarray}}
\newcommand{\eeqa}{\end{eqnarray}}
\newcommand{\non}{\nonumber}
\newcommand{\lab}{\label}

\newcommand{\bra}{\langle}
\newcommand{\ket}{\rangle}

\newfont{\bg}{cmr10 scaled\magstep4}

\newcommand{\bigzerou}{\smash{\lower1.7ex\hbox{\bg 0}}}
\newcommand{\lw}[1]{\smash{\lower2.0ex\hbox{#1}}}
\newcommand{\mapright}[1]{\smash{\mathop{\hbox to 1cm{\rightarrowfill}}\limits^{#1}}}

\newcommand{\rightcircular}{\supset\kern-1.3em\lower0.67ex\hbox{$\triangleleft$}}
\newcommand{\leftcircular}{\subset\kern-0.55em\lower0.67ex\hbox{$\triangleright$}}
\newcommand{\smallrightcircular}{\supset\kern-0.75em\lower0.46ex\hbox{\scriptsize$\triangleleft\,\,\,\,\,$}}
\newcommand{\smallleftcircular}{\subset\kern-0.17em\lower0.46ex\hbox{\scriptsize$\triangleright$}}

\begin{document}

\title{A method of enciphering quantum states}

\author{Hiroo Azuma\thanks{Present address: Centre for Quantum Computation,
Clarendon Laboratory,
Parks Road, Oxford OX1 3PU, UK.
E-mail: hiroo.azuma@qubit.org}
\\
{\small Mathematical Engineering Division,
Canon Research Center,}\\
{\small 5-1, Morinosato-Wakamiya, Atsugi-shi,
Kanagawa, 243-0193, Japan}\\
{\small E-mail: hiroo@crc.canon.co.jp}
\and
Masashi Ban\\
{\small Advanced Research Laboratory,
Hitachi Ltd.,}\\
{\small Hatoyama, Saitama, 350-0395, Japan}\\
{\small E-mail: m-ban@harl.hitachi.co.jp}}

\date{December 2, 2000}

\maketitle

\begin{abstract}
In this paper,
we propose a method of enciphering quantum states
of two-state systems (qubits)
for sending them in secrecy
without entangled qubits
shared by two legitimate users (Alice and Bob).
This method has the following two properties.
First, even if an eavesdropper (Eve) steals qubits,
she can extract information from them
with certain probability at most.
Second, Alice and Bob can confirm
that the qubits are transmitted between them correctly
by measuring a signature.
If Eve measures $m$ qubits one by one
from $n$ enciphered qubits
and sends alternative ones (the Intercept/Resend attack),
a probability that Alice and Bob do not notice Eve's action is
equal to $(3/4)^{m}$ or less.
Passwords for decryption and the signature are given
by classical binary strings
and they are disclosed through a public channel.
Enciphering classical information by this method is
equivalent to the one-time pad method
with distributing a classical key
(random binary string) by the BB84 protocol.
If Eve takes away qubits,
Alice and Bob lose the original quantum information.
If we apply our method to a state in iteration,
Eve's success probability decreases exponentially.
We cannot examine security against the case
that Eve makes an attack with using entanglement.
This remains to be solved in the future.
\end{abstract}

\section{Introduction}\lab{introduction}

Since considerable progress was made
in quantum information and computation theory,
many researchers have been studying
to realize the information processing that we have never had
by using quantum mechanics\cite{shor-deutsch-jozsa}.
At the same time,
it has been studying to apply the uncertainty principle,
the quantum no-cloning theorem,
and entanglement between quantum systems
to cryptography\cite{QNoCloning}.
The BB84 protocol is considered to be an effective method
for the key distribution.
By combining it with the one-time pad method,
we obtain a highly secure
cryptography\cite{BB84org}\cite{bennett-BB84exp}\cite{bennett-brassard-ekert}.
On the other hand,
the quantum teleportation is considered to be an excellent method
for sending arbitrary quantum
states between two
parties\cite{QTeleportOrg}\cite{QTeleportExp-bouwmeester-furusawa}.

The BB84 protocol is used for a secure distribution
of a classical key
(binary string) to two legitimate users (Alice and Bob).
Choosing
a basis vector at random
from four basis vectors, the rectilinear basis $\{|0\ket,|1\ket\}$
and the circular basis $\{(1/\sqrt{2})(|0\ket\pm|1\ket)\}$,
as a state of a photon (a two-state system or a qubit),
Alice sends it to Bob.
Bob measures a transmitted photon
in an orthonormal basis that he chooses
from two bases (rectilinear and circular)
at random and independently of Alice.

Not being consistent with each other,
the rectilinear basis and the circular basis
are called conjugate bases.
A result of measurement with a wrong basis is random.
If an eavesdropper (Eve) steals a photon from the channel,
measures it in a basis chosen at random,
and sends an alternative one,
Alice and Bob find inconsistency with probability of $1/4$ or more
and notice Eve's eavesdropping.
In this way, by using the uncertainty principle,
the BB84 protocol reveals Eve's illegal act.

A.~K.~Ekert
proposed another protocol
for distributing a classical key
by transmitting pairs of qubits
in EPR states, $|\Psi^{-}\ket=(1/\sqrt{2})(|01\ket-|10\ket)$,
apart to Alice and Bob from a source\cite{Ekert91}.
They detect Eve by Bell's theorem.
Considering a simplified protocol of A.~K.~Ekert's,
C.~H.~Bennett et al. showed it was equivalent
to BB84\cite{BBM92_VSEkert91}.
From these successive works,
it has been recognized that
we do not need to use the entanglement
for distributing a classical key by quantum mechanics.
(But, by combining entanglement purification protocol
with Ekert's protocol,
we can distribute a classical key
in high secrecy\cite{EntanglementPurification}.)

The quantum teleportation is used for transmitting
an arbitrary state from Alice to Bob.
They share an EPR-pair of qubits beforehand.
Alice carries out the Bell-measurement
on both a one-qubit state $|\psi\ket$
that she wants to send
and her qubit of the EPR-pair.
Receiving a result of her measurement,
Bob can construct $|\psi\ket$ from his qubit of the EPR-pair.
A characteristic of this method is that
classical information and non-classical information of $|\psi\ket$
are divided perfectly
and only the classical information is sent through the public channel.
If they share the EPR-pair correctly,
Eve can neither eavesdrop on the state nor destroy it in principle.

These methods are related to the quantum no-cloning theorem.
It tells us
there is no unitary transformation that makes
accurate clones of arbitrary quantum states\cite{QNoCloning}.
In the BB84 protocol,
it gives an effect as follows.
Not knowing which basis is chosen
for a qubit (photon) that she extracts from the quantum channel,
rectilinear or circular,
Eve cannot make a clone of the qubit and keep it.
What she can do is only measuring the qubit
in a proper basis
and sending an alternative one that depends
on a result of the measurement to Bob.
In the quantum teleportation, the following thing is important.
Because Alice can neither measure $|\psi\ket$
without disturbance
nor make an accurate clone of it,
she cannot extract information from $|\psi\ket$ at all.
During the whole process,
Alice and Bob have no knowledge about $|\psi\ket$.

In the quantum teleportation,
Alice and Bob have to share an EPR-pair of qubits
beforehand.
After being emitted by a source,
this pair flies towards them apart through a quantum channel.
Therefore, for example,
if Eve takes away the qubit that Bob is supposed to have
and sends an alternative one to him,
she succeeds in eavesdropping.
To avoid such a trouble,
Alice and Bob need to share a lot of EPR-pairs
and to purify them\cite{EntanglementPurification}.

In this paper,
we consider a method for enciphering arbitrary quantum states
for sending them in secrecy
without entangled qubits shared by Alice and Bob beforehand.
In our method,
there are two points as follows. (See Figure~\ref{protocol_1}.)

First,
even if Eve takes away qubits,
she can extract quantum information from them
with certain probability at most.
(If Eve measures $m$ qubits one by one
from $n$ enciphered qubits
and sends alternative ones,
a probability that Alice and Bob do not notice Eve's act is
equal to $(3/4)^{m}$ or less.
We assume Eve makes only
the Intercept/Resend attack\cite{bennett-BB84exp}.)
Alice applies a unitary operator $U_{i}$
which is chosen at random from a set of operators
${\cal M}=\{U_{j}\}$
to an arbitrary $n$-qubit state $|\Psi\ket$ that she wants to send
in secrecy.
The subscript $i$ of $U_{i}$ is a password for decryption.
Not knowing which operator is chosen from ${\cal M}$,
Eve regards the enciphered state
as a mixed state of
$U_{j}|\Psi\ket$ for all $U_{j}\in{\cal M}$
with equal probability.
If Alice prepares ${\cal M}$
so that
the density operator of the mixed state may be
in proportion to the identity operator $\mbox{\boldmath $I$}$,
Eve cannot extract the information of $|\Psi\ket$ at all
without the password $i$.
The reason of this is that
even if Eve puts auxiliary qubits on
the density operator $\rho=(1/2^{n})\mbox{\boldmath $I$}$,
applies unitary transformations to it,
or measures it,
she cannot extract $|\Psi\ket$.
After confirming that the quantum state is transmitted correctly,
Alice releases the password $i$ in our protocol.
Therefore,
to extract information from $|\Psi\ket$,
Eve has to eavesdrop without disturbing
Alice and Bob's certification process.
(This technique has been also discussed
by P.~O.~Boykin and V.~Roychowdhury,
and M.~Mosca et al.\cite{Boykin-Mosca}.
They have shown the following result.
When we define ${\cal M}$
as a set of tensor products of the Pauli matrices,
the number of the operators $\{U_{i}\}$ gets minimum
and the subscript $i$ is represented by a $2n$-bit string.)

Second,
Alice and Bob can confirm that
a quantum state received by Bob is a genuine one sent by Alice.
Not having knowledge about the $n$-qubit state
$|\Psi\ket_{Q}$ at all,
they do not notice Eve replace the genuine qubits with alternative ones.
Therefore,
they need to confirm that
the qubits Bob receives are genuine.
(It seems like authentication of the identity of a correspondent
on networks.)
In our method,
after putting an $n$-qubit state $|a\ket_{S}$
($a\in\{0,1\}^{n}$)
that represents her signature
on $U_{i}^{Q}|\Psi\ket_{Q}$,
Alice makes entanglement between qubits of each pair
in the system $Q$ and $S$.
Here,
we call the quantum system which represents the transmitted
information $Q$
and
the quantum system which represents the signature $S$.
Then,
to forbid Eve to make clones of qubits,
Alice applies an operator chosen at random
from ${\cal L}=\{\mbox{\boldmath $I$}, H, \sigma_{x}, H\sigma_{x}\}$
to each qubit,
where
$H$ is called the Hadamard transformation
and
it causes
$|0\ket\rightarrow(1/\sqrt{2})(|0\ket+|1\ket)$,
$|1\ket\rightarrow(1/\sqrt{2})(|0\ket-|1\ket)$,
and
$\sigma_{x}$ is one of the Pauli matrices
and
it causes
$|0\ket\rightarrow|1\ket$,
$|1\ket\rightarrow|0\ket$.
Hence,
quantum information of each qubit is encoded
in a basis chosen at random from two conjugate bases
(rectilinear and circular).
Therefore, if Eve does anything on the qubits,
Alice and Bob can find inconsistency
and detect Eve
with certain probability.
This is essentially the same technique
used in BB84.
In our method,
certification of a correspondent
and detection of Eve
are done at the same time.
The second password for decryption is
which operators are chosen from ${\cal L}$.

Because the passwords and the signature
represented by classical binary strings
are transmitted by the public channel,
Eve also knows them.
If $|\Psi\ket_{Q}$ represents classical information
(a product state of $|0\ket$ and $|1\ket$),
our protocol is equivalent to the one-time pad method
with classical key distribution by BB84.

If Alice and Bob apply our enciphering method
to a state in iteration,
a probability that Eve gets quantum information
with fidelity of $1$ decreases exponentially.
(If they encipher a one-qubit state with $N$ qubits,
the probability is given by $(3/4)^{N/2}$.)
We can regard it as privacy amplification process.
We cannot examine security against the case
that Eve makes an attack with using entanglement.
This remains to be solved in the future.

This paper is organized as follows.
In Sec.~\ref{sec-cryptize-qstate},
we explain how Alice and Bob forbid Eve
to extract the original quantum information
and how they confirm the qubits are transmitted correctly.
In Sec.~\ref{q-crypto-protocol},
we explain the whole protocol and
discuss how Bob confirms that he receives qubits.
In Sec.~\ref{Eve-strategy},
we discuss security of our protocol
against Eve's Intercept/Resend attack on each qubit.
In Sec.~\ref{Privacy-Amplification},
we discuss privacy amplification process.
In Sec.~\ref{CONCLUSION}, we give a brief discussion.

\section{Enciphering quantum states}\lab{sec-cryptize-qstate}

In this section,
we explain how Alice and Bob
forbid Eve to extract the original quantum information
and how they confirm
the qubits are transmitted correctly.

First,
we consider
transforming an arbitrary quantum state
so that Eve cannot recover an original quantum state.
For simplicity,
we consider an arbitrary one-qubit state for a while
and we describe its density operator as $\rho$
defined on a two-dimensional Hilbert space ${\cal H}_{2}$.
We assume Alice wants to send $\rho$ to Bob in secrecy.
She does not know $\rho$ at all,
because
her partial measurement destroys it.

Alice prepares a set of operators,
\[
{\cal M}=\{\sigma_{j} : j=0,x,y,z\},
\]
where
$\sigma_{0}=\mbox{\boldmath $I$}$
(the identity operator)
and
$\{\sigma_{x}, \sigma_{y},\sigma_{z} \}$
are the Pauli matrices.
Taking
\[
|0\ket=\left(\begin{array}{c}
1 \\
0 \\
\end{array}\right),\quad
|1\ket=\left(\begin{array}{c}
0 \\
1 \\
\end{array}\right),
\]
for an orthonormal basis, we write them as
\[
\mbox{\boldmath $I$}
=\left(\begin{array}{cc}
1 & 0 \\
0 & 1 \\
\end{array}\right),\quad
\sigma_{x}=\left(\begin{array}{cc}
0 & 1 \\
1 & 0 \\
\end{array}\right),\quad
\sigma_{y}=\left(\begin{array}{cc}
0 & -i \\
i & 0 \\
\end{array}\right),\quad
\sigma_{z}=\left(\begin{array}{cc}
1 & 0 \\
0 & -1 \\
\end{array}\right).
\]
${\cal M}$
may be disclosed in public.
Choosing an operator $\sigma_{i}$
from ${\cal M}$ at random,
Alice carries out the following unitary transformation,
\[
\rho \rightarrow \sigma_{i}\rho \sigma_{i}^{\dagger}.
\]
She keeps the subscript $i$ secret as a password
and never tells it to anyone.
Because the subscript takes a value from
$\{0,x,y,z\}$,
the password can be represented by $2$-bit classical information.

Not knowing which transformation Alice applies to the qubit,
Eve has to regard the state as
\[
\rho'=\frac{1}{4}\sum_{j=0,x,y,z}
\sigma_{j}\rho\sigma_{j}^{\dagger}.
\]
Generally,
the density operator $\rho$ satisfies
$\rho^{\dagger}=\rho$,
$\mbox{Tr}\rho=1$,
$0 \leq \lambda_{1}\leq 1$,
$0 \leq \lambda_{2}\leq 1$,
and
$\lambda_{1}+\lambda_{2}=1$
where
$\lambda_{1}$, $\lambda_{2}$ are eigenvalues of $\rho$.
Hence,
we can describe an arbitrary $\rho$ as
\[
\rho=\frac{1}{2}
(\mbox{\boldmath $I$}+
\mbox{\boldmath $a$}\cdot \mbox{\boldmath $\sigma$}),
\]
where $\mbox{\boldmath $a$}=(a_{1},a_{2},a_{3})$
is a three-component real vector
and
$0\leq\sum_{k=1}^{3}a_{k}^{2}\leq 1$.
Because
\beq
\rho'
=
\frac{1}{2}\mbox{\boldmath $I$}+
\frac{1}{8}\sum_{j=0,x,y,z}
\mbox{\boldmath $a$}\cdot
\sigma_{j}
\mbox{\boldmath $\sigma$}
\sigma_{j}^{\dagger},
\lab{1-qubit-cry-mix}
\eeq
and
\[
\sigma_{j}\sigma_{k}\sigma_{j}=
\left\{
\begin{array}{ll}
\sigma_{k} & \mbox{$j=0$ or $j=k$} \\
-\sigma_{k} & \mbox{$j\neq k$ and $j,k \in\{x,y,z \}$}
\end{array}
\right .,
\]
there is no contribution from the second term of Eq.~(\ref{1-qubit-cry-mix}).
We obtain
\[
\rho'=\frac{1}{2}\mbox{\boldmath $I$}.
\]
Therefore,
even if Eve takes away $\rho'$,
she cannot extract information from it at all,
because she does not know the password $i$.

An arbitrary $n$-qubit density operator $\rho_{n}$ is given by
\beq
\rho_{n}=\frac{1}{2^{n}}
(\mbox{\boldmath $I$}+
\sum_{
\mbox{\scriptsize \boldmath $k$}\in\{0,x,y,z\}^{n},
\mbox{\scriptsize \boldmath $k$}\neq(0,\cdots,0)
}
a_{
\mbox{\scriptsize \boldmath $k$}
}
U_{
\mbox{\scriptsize \boldmath $k$}
}
),
\lab{n-qubit-density-matrix-general}
\eeq
where
\[
U_{
\mbox{\scriptsize \boldmath $k$}
}
=\sigma_{k_{1}}\otimes \cdots \otimes \sigma_{k_{n}},
\]
and
$a_{
\mbox{\scriptsize \boldmath $k$}
}
$
($\mbox{\boldmath $k$}\in\{0,x,y,z \}^{n}$,
$\mbox{\boldmath $k$}\neq(0,\cdots,0)$)
is real.
(In Eq.~(\ref{n-qubit-density-matrix-general}),
$\mbox{\boldmath $I$}$ represents the identity operator
for $n$-qubit states.)
Choosing an operator
$U_{
\mbox{\scriptsize \boldmath $i$}
}$
from
\[
{\cal M}_{n}=\{
U_{
\mbox{\scriptsize \boldmath $k$}
}
:
U_{
\mbox{\scriptsize \boldmath $k$}
}
=\sigma_{k_{1}}\otimes \cdots \otimes \sigma_{k_{n}},
\mbox{\boldmath $k$}
\in\{0,x,y,z\}^{n}
\},
\]
at random,
Alice applies it to $\rho_{n}$
for encryption as
$
\rho_{n}
\rightarrow
U_{
\mbox{\scriptsize \boldmath $i$}
}
\rho_{n}
U_{
\mbox{\scriptsize \boldmath $i$}
}^{\dagger}
$.
If Eve takes away the density operator given by
\beqa
\rho_{n}'
&=&
\frac{1}{4^{n}}
\sum_{\mbox{\scriptsize \boldmath $j$}\in\{0,x,y,z\}^{n}}
U_{
\mbox{\scriptsize \boldmath $j$}
}
\rho_{n}
U_{
\mbox{\scriptsize \boldmath $j$}
}^{\dagger} \non \\
&=&
\frac{1}{2^{n}}\mbox{\boldmath $I$}
+
\frac{1}{4^{n}\cdot 2^{n}}
\sum_{
\mbox{\scriptsize \boldmath $j$},
\mbox{\scriptsize \boldmath $k$}
\in\{0,x,y,z\}^{n},
\mbox{\scriptsize \boldmath $k$}\neq(0,\cdots,0)
}
a_{
\mbox{\scriptsize \boldmath $k$}
}
U_{
\mbox{\scriptsize \boldmath $j$}
}
U_{
\mbox{\scriptsize \boldmath $k$}
}
U_{
\mbox{\scriptsize \boldmath $j$}
}^{\dagger} \non \\
&=&
\frac{1}{2^{n}}\mbox{\boldmath $I$},
\lab{n-qubit-cry-mix}
\eeqa
she cannot extract information from $\rho_{n}'$ at all.
Even if she puts auxiliary systems on $\rho_{n}'$,
applies unitary transformations to it,
or carries out measurements,
she cannot obtain $\rho_{n}$.
The password of $\mbox{\boldmath $i$}$
is given by a $2n$-bit string.
(This technique is also discussed
by P.~O.~Boykin and V.~Roychowdhury,
and M.~Mosca et al.
as mentioned in Sec.~\ref{introduction}\cite{Boykin-Mosca}.)

Next,
we explain how Alice and Bob confirm that
the qubits are transmitted between them correctly.
If Eve takes away $\rho_{n}'$
and sends an alternative state $\tilde{\rho}_{n}$ to Bob,
he carries out the inverse operation
on the state that he receives as
$\tilde{\rho}_{n}
\rightarrow
U_{
\mbox{\scriptsize \boldmath $i$}
}
\tilde{\rho}_{n}
U_{
\mbox{\scriptsize \boldmath $i$}
}^{\dagger}$.
Not having knowledge about the original $\rho_{n}$ at all,
Alice and Bob do not notice what Bob gets is a fake.

To avoid this trouble,
they put a signature on $\rho_{n}$.
Reading it,
they can confirm that Bob receives the state transmitted
from Alice correctly.
It is important that Eve cannot
change the signature.
For simplicity,
we describe the state as a ket vector
$\forall |\Psi\ket_{Q}\in{\cal H}_{2}^{n}$
instead of the density operator $\rho_{n}$ for a while.
If the $n$-qubit state is given by a mixed state,
we can give a similar discussion.
We call the system that represents the quantum information $Q$
and the system that represents the signature $S$.

\begin{figure}
\caption{The second encryption by Alice.}
\begin{center}
\includegraphics[scale=0.9]{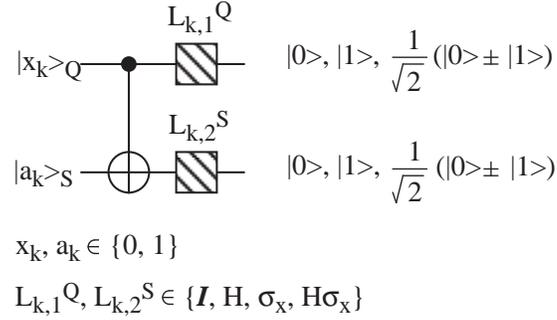}
\end{center}
\lab{2nd-criptization1}
\end{figure}

Preparing an $n$-bit random string
$\forall\mbox{\boldmath $a$}=(a_{1},\cdots,a_{n})\in\{0,1\}^{n}$
as her signature,
Alice attaches a qubit $|a_{k}\ket_{S}$
to the $k$th qubit of
\beq
U_{
\mbox{\scriptsize \boldmath $i$}
}^{Q}|\Psi\ket_{Q}
=\sum_{\mbox{\scriptsize\boldmath $x$} \in\{0,1\}^{n}}
c_{\mbox{\scriptsize\boldmath $x$}}
|x_{1}\ket\cdots|x_{n}\ket
\in {\cal H}_{2}^{n}.
\lab{n-qbt-genrl-entg-state}
\eeq
Applying the controlled-NOT (C-NOT) gate
to the $k$th pair
as Figure~\ref{2nd-criptization1},
she obtains
\[
|x_{k}\ket_{Q}|a_{k}\ket_{S}
\rightarrow
|x_{k}\ket_{Q}|a_{k}\oplus x_{k}\mbox{ mod 2}\ket_{S}
\quad\quad
\mbox{for $k=1,\cdots,n$}.
\]
Choosing
$L_{k,1}^{Q}, L_{k,2}^{S}\in{\cal L}
=\{\mbox{\boldmath $I$}, H, \sigma_{x}, H\sigma_{x}\}$
at random,
she applies them to the qubits of the pair,
where
\[
H=\frac{1}{\sqrt{2}}
\left(
\begin{array}{cc}
1 & 1\\
1 & -1 \\
\end{array}
\right).
\]
Alice repeats this operation on the $k=1,\cdots,n$th qubit,
and we write the whole transformation as
$V_{\mbox{\scriptsize \boldmath $\alpha$}}^{QS}$.
$\mbox{\boldmath $\alpha$}$
represents
which operators are chosen from ${\cal L}$.
The second password $\mbox{\boldmath $\alpha$}$
is given by a classical $4n$-bit string.

\begin{figure}
\caption{Typical quantum gates.}
\begin{center}
\includegraphics[scale=0.9]{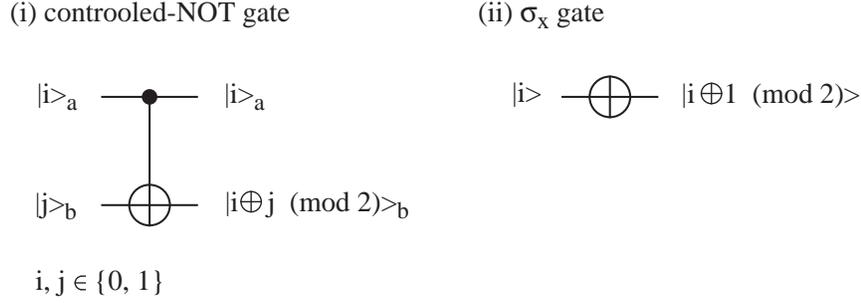}
\end{center}
\lab{element-q-gates}
\end{figure}

We often describe successive operations on qubits as
a network like Figure~\ref{2nd-criptization1}.
A horizontal line represents a qubit
and time proceeds from left to right.
Figure~\ref{element-q-gates} shows examples of
unitary transformations applied to qubits.
Figure~\ref{element-q-gates}.~i and ii. represent
the C-NOT gate
and $\sigma_{x}$ each\cite{feynman-barenco-bennett}.

The operations that we have discussed are
summarized as follows,
\[
\begin{array}{lcll}
|\Psi\ket_{Q}
&\rightarrow&
U_{
\mbox{\scriptsize \boldmath $i$}
}^{Q}
|\Psi\ket_{Q}
&
\mbox{(1st password {$\mbox{\boldmath $i$}$})} \\
&\rightarrow&
|\mbox{\boldmath $a$}\ket_{S}
\otimes
U_{
\mbox{\scriptsize \boldmath $i$}
}^{Q}
|\Psi\ket_{Q}
&
\mbox{(signature {\boldmath $a$})} \non \\
&\rightarrow&
V_{\mbox{\scriptsize \boldmath $\alpha$}}^{QS}
[|\mbox{\boldmath $a$}\ket_{S}
\otimes
U_{
\mbox{\scriptsize \boldmath $i$}
}^{Q}
|\Psi\ket_{Q}]
&
\mbox{(2nd password $\mbox{\boldmath $\alpha$}$)}. \\
\end{array}
\]
Double encryption
prevents Eve from using $|\mbox{\boldmath $a$}\ket_{S}$
ill.
If $V_{\mbox{\scriptsize \boldmath $\alpha$}}^{QS}$ is not applied
to the state,
Eve may take away all of the qubits,
keep
$U_{
\mbox{\scriptsize \boldmath $i$}
}^{Q}
|\Psi\ket_{Q}
$,
and send
a fake of
$|\mbox{\boldmath $a$}\ket_{S}\otimes|\tilde{\Psi}\ket_{Q}$
to Bob.

If Eve does anything on the $k$th qubit of $Q$,
the signature of $|a_{k}\ket_{S}$ is destroyed
and
Bob fails in the certification process
with certain probability or more.
This is caused by the facts
that
conjugate bases chosen at random
represent
the systems $Q$ and $S$,
and
there is entanglement between $Q$ and $S$.
If she does anything on the $k$th qubit of $S$,
we can get a similar result.
We estimate a probability
that Alice and Bob notice Eve's illegal act
in Sec.~\ref{Eve-strategy}.

\section{The protocol for secure transmission}\lab{q-crypto-protocol}

We consider a protocol
for transmitting an arbitrary quantum state
$\forall|\Psi\ket_{Q}\in{\cal H}_{2}^{n}$
from Alice to Bob
in secrecy
against Eve's eavesdropping
by using the encryption method discussed in the previous section.
Alice and Bob do not have knowledge about $|\Psi\ket_{Q}$ at all.
Both of them can use the following two channels.
\begin{itemize}
\item
The classical channel:
It transmits classical binary strings in public.
Eve can make accurate copies of them,
but she cannot alter them.
\item
The quantum channel:
It transmits sequences of qubits (quantum information).
Eve can interact with them,
but she cannot make accurate copies of them.
\end{itemize}

\begin{figure}
\caption{Secure transmission between Alice and Bob.}
\begin{center}
\includegraphics[scale=0.9]{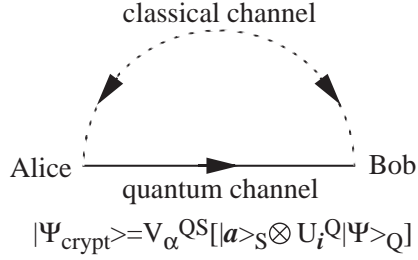}
\end{center}
\lab{protocol_1}
\end{figure}

Alice sends qubits to Bob according to the following protocol.
(See Figure~\ref{protocol_1}.)
\begin{enumerate}
\item
Alice sends
$
|\Psi_{
\mbox{\scriptsize crypt}
}\ket
\equiv
V_{\mbox{\scriptsize \boldmath $\alpha$}}^{QS}
[|\mbox{\boldmath $a$}\ket_{S}
\otimes
U_{
\mbox{\scriptsize \boldmath $i$}
}^{Q}
|\Psi\ket_{Q}]$
of $2n$ qubits
to Bob through the quantum channel.
\item
Receiving $2n$ qubits,
Bob breaks off the quantum channel
and reports arrival of them
to Alice through the classical channel.
\item
Receiving the report from Bob,
Alice tells Bob what transformation
$V_{\mbox{\scriptsize \boldmath $\alpha$}}^{QS}$
is through the classical channel.
(She discloses the $4n$-bit password $\mbox{\boldmath $\alpha$}$.)
\item
Applying $V_{\mbox{\scriptsize \boldmath $\alpha$}}^{QS\dagger}$
to the state that he has received and measuring the signature,
Bob tells Alice a result of the measurement (an $n$-bit string)
through the classical channel.
\item
Receiving the $n$-bit string from Bob,
Alice examines
whether it coincides with the signature
$\mbox{\boldmath $a$}$ or not.
If it coincides with her original signature,
she tells Bob what transformation
$U_{
\mbox{\scriptsize \boldmath $i$}
}^{Q}$
is through the classical channel.
(She discloses the $2n$-bit password $\mbox{\boldmath $i$}$.)
If it does not coincide,
she concludes Eve has eavesdropped on qubits
and stops the protocol.
\item
Bob applies
$U_{
\mbox{\scriptsize \boldmath $i$}
}^{Q\dagger}$
to the state that he has
and obtains the original state
$|\Psi\ket_{Q}$.
\end{enumerate}

\begin{figure}
\caption{Eve's strategy for eavesdropping.}
\begin{center}
\includegraphics[scale=0.9]{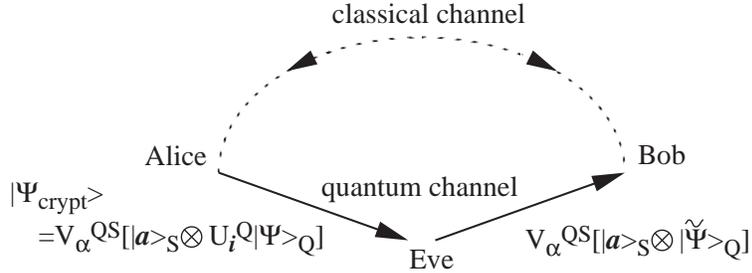}
\end{center}
\lab{protocol_2}
\end{figure}

In this protocol,
it is important
that
Bob confirms the arrival of $2n$ qubits
($|\Psi_{
\mbox{\scriptsize crypt}
}\ket$
or Eve's fake)
and breaks off the quantum channel
at the second step.
To understand the reason of this,
we assume the following case.
(See Figure~\ref{protocol_2}.)
Although $|\Psi_{
\mbox{\scriptsize crypt}
}\ket$
is still halfway on the channel,
Bob reports the arrival of qubits to Alice
by mistake,
and
she discloses $V_{\mbox{\scriptsize \boldmath $\alpha$}}^{QS}$
through the classical channel.
Eve may take away
$|\Psi_{
\mbox{\scriptsize crypt}
}\ket$,
apply $V_{\mbox{\scriptsize \boldmath $\alpha$}}^{QS\dagger}$
to it,
and obtain
$|\mbox{\boldmath $a$}\ket_{S}
\otimes
U_{
\mbox{\scriptsize \boldmath $i$}
}
|\Psi\ket_{Q}$
before Bob receives qubits.
Eve can keep $U_{
\mbox{\scriptsize \boldmath $i$}
}
|\Psi\ket_{Q}$,
combine a false state $|\tilde{\Psi}\ket_{Q}$
with
$|\mbox{\boldmath $a$}\ket_{S}$,
and send
$V_{\mbox{\scriptsize \boldmath $\alpha$}}^{QS}
[|\mbox{\boldmath $a$}\ket_{S}
\otimes
|\tilde{\Psi}\ket_{Q}]$
to Bob.
If the quantum channel is still open,
Bob receives
$V_{\mbox{\scriptsize \boldmath $\alpha$}}^{QS}
[|\mbox{\boldmath $a$}\ket_{S}
\otimes
|\tilde{\Psi}\ket_{Q}]$.
Because the signature is correct,
Alice and Bob cannot notice Eve's illegal act.
Alice discloses $U_{
\mbox{\scriptsize \boldmath $i$}
}^{Q}
$ in public
and finally
Eve gets $|\Psi\ket_{Q}$.

\begin{figure}
\caption{The photon counting
measurement with nonlinear optical devices.}
\begin{center}
\includegraphics[scale=0.9]{cavityQPG2e.epsf}
\end{center}
\lab{cavityQPG2}
\end{figure}

To avoid this trouble,
Bob needs to verify that a batch of qubits
($|\Psi_{
\mbox{\scriptsize crypt}
}\ket$ or Eve's fake) has arrived.
For example,
it is good for Bob to take the following method.

We construct a qubit from a pair of optical paths (modes)
that are represented by $x$ and $y$
in Figure~\ref{cavityQPG2}.~iii\cite{ChuangYamamoto}.
We describe a state that there is no photon on a mode as $|0\ket$
and a state
that there is a photon on a mode as $|1\ket$.
Writing a state that no photon is on the mode $x$
and one photon is on the mode $y$
as
$|0\ket_{x}\otimes|1\ket_{y}=|01\ket$,
we regard $|01\ket$
as
logical $|\bar{0}\ket$.
We regard $|1\ket_{x}|0\ket_{y}$
as logical $|\bar{1}\ket$ similarly.
Hence, we can write an arbitrary state of a qubit as
\[
|\psi\ket=\alpha|01\ket+\beta|10\ket
=\alpha|\bar{0}\ket+\beta|\bar{1}\ket
\quad\quad
\mbox{for $|\alpha|^{2}+|\beta|^{2}=1$.}
\]

We assume
Eve puts a 50-50 beamsplitter halfway on the quantum channel
to take away photons.
A state of a photon is a superposition of
a state that it is on the side of Bob
and
a state that it is on the side of Eve
with amplitude $1/\sqrt{2}$ each,
\[
\frac{1}{\sqrt{2}}
|\mbox{photon on the side of Bob}\ket
+\frac{1}{\sqrt{2}}
|\mbox{photon on the side of Eve}\ket.
\]
To examine whether the qubit of $|\psi\ket$ has come on his own side or not,
Bob prepares another auxiliary photon
and applies nonlinear interaction
between the logical photon on the mode $x$ or $y$ and the auxiliary one.
On an optical system of Figure~\ref{cavityQPG2}.~iii,
if the photon counter $D_{a}$ detects the auxiliary photon,
the logical photon is projected
into the state that it is on the side of Bob.
On the other hand,
$D_{b}$'s detection projects the logical photon
into the state that it is on the side of Eve.

In Figure~\ref{cavityQPG2}.~iii,
there are
beamsplitters $B$ which apply $SU(2)$ transformations
to logical kets
$\{|\bar{0}\ket,|\bar{1}\ket\}$
as Figure~\ref{cavityQPG2}.~i,
and Kerr-type devices $K$ which induce nonlinear interactions
between two incoming photons as Figure~\ref{cavityQPG2}.~ii.
The device $K$ shifts a phase of a wave function
by $\pi$
only if a pair of photons comes into it.
(Q.~A.~Turchette et al.
succeeds in shifting the phase
by $\Delta\sim 16^{\circ}$\cite{QPPhaseGateExp}.)

\begin{figure}
\caption{A quantum network for photon counting.}
\begin{center}
\includegraphics[scale=0.9]{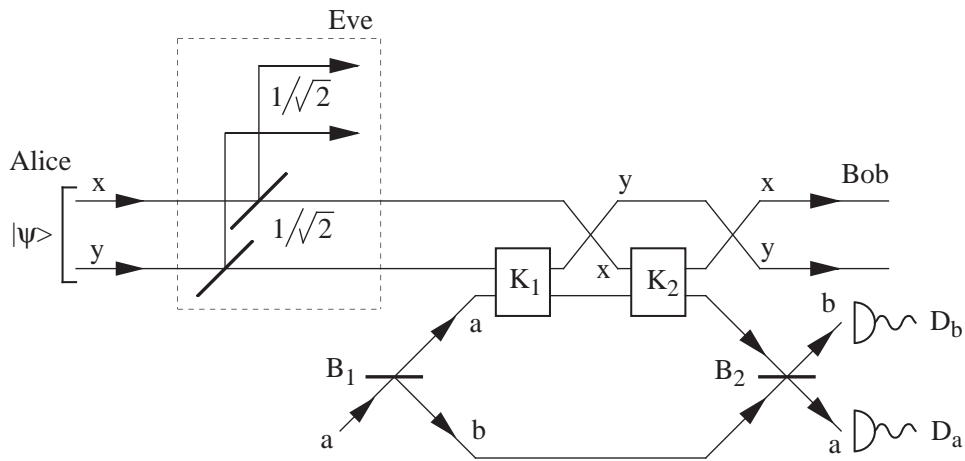}
\end{center}
\lab{phtcount_net1}
\end{figure}

To clarify the operation of Figure~\ref{cavityQPG2}.~iii,
we describe it by a network of quantum gates
in Figure~\ref{phtcount_net1}\cite{Gottesman}.
Assuming the first and the second qubits are
in an arbitrary entangled state $|\Psi\ket_{Q}$,
we examine whether the first qubit exists or not
by measuring an auxiliary qubit system $A$.
When we write the whole system as
\[
|\Psi\ket_{Q}|0\ket_{A}=\sum_{i,j\in\{0,1\}}
c_{ij}|i\ket_{2}|j\ket_{1}|0\ket_{A},
\]
$|\Psi\ket_{Q}|0\ket_{A}$
is transformed as follows in Figure~\ref{phtcount_net1},
\beqa
\lefteqn{|\Psi\ket_{Q}|0\ket_{A}
=
\sum_{i}
(c_{i0}|i\ket_{2}|0\ket_{1}
+c_{i1}|i\ket_{2}|1\ket_{1})|0\ket_{A}} \non \\
&\mapright{\mbox{\scriptsize step 1}}&
\sum_{i}
(c_{i0}|i\ket_{2}|1\ket_{1}|0\ket_{A}
+c_{i1}|i\ket_{2}|0\ket_{1}|1\ket_{A}) \non \\
&\mapright{\mbox{\scriptsize step 2}}&
\sum_{i}
(c_{i0}|i\ket_{2}|0\ket_{1}
+c_{i1}|i\ket_{2}|1\ket_{1})|1\ket_{A} \non \\
&&=|\Psi\ket_{Q}|1\ket_{A}. \non
\eeqa

Therefore,
measuring an auxiliary system
as shown in Figures~\ref{cavityQPG2}
and \ref{phtcount_net1}
for each channel,
Bob can examine whether all of the qubits are arrived or not.

\section{Security against eavesdropping}\lab{Eve-strategy}

It is difficult to consider all strategies Eve may take.
In this section,
we assume Eve to make only the Intercept/Resend attack.
Eve measures each transmitted qubit with a proper basis independently
and sends an alternative one
according to the result of the measurement\cite{bennett-BB84exp}.
We pay attention to the following fact.
Eve cannot extract information about $|\Psi\ket_{Q}$ at all
without getting the first password $\mbox{\boldmath $i$}$
of $U_{
\mbox{\scriptsize \boldmath $i$}
}^{Q}|\Psi\ket_{Q}$,
because the enciphered density operator is
in proportion to $\mbox{\boldmath $I$}$ for her.
Therefore,
Eve needs to keep her illegal action
secret from Alice and Bob during their authentication process
so that
Alice may disclose the first password $\mbox{\boldmath $i$}$.
In this section,
we estimate a probability
that Alice and Bob fail to notice Eve's illegal act.

For simplicity,
we  assume that
$|\Psi\ket_{Q}$ is an arbitrary $n$-qubit product state
for a while.
At the first encryption,
Alice applies
$\forall U_{\mbox{\scriptsize \boldmath $i$}}^{Q}
=\sigma_{i_{1}}\otimes\cdots\otimes\sigma_{i_{n}}$
to $|\Psi\ket_{Q}=|\psi_{1}\ket\otimes\cdots\otimes|\psi_{n}\ket$.
Hence,
$U_{\mbox{\scriptsize \boldmath $i$}}^{Q}|\Psi\ket_{Q}$
is also a product state
and
we may treat each qubit independently.

\begin{figure}
\caption{Eve's Intercept/Resend attack on the system $S$.}
\begin{center}
\includegraphics[scale=0.9]{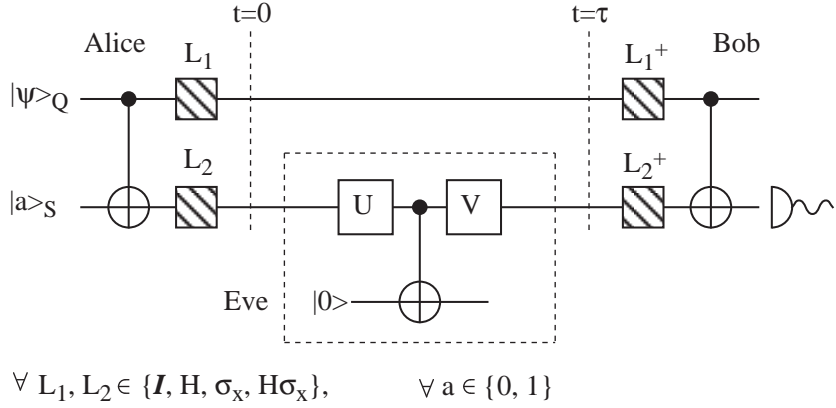}
\end{center}
\lab{Eve-Strategy1}
\end{figure}

We write the $k$th qubit ($\forall k\in\{1,\cdots,n\}$) of
$U_{
\mbox{\scriptsize \boldmath $i$}
}^{Q}|\Psi\ket_{Q}$
as
$|\psi\ket_{Q}=\alpha|0\ket_{Q}+\beta|1\ket_{Q}$
and the $k$th qubit of the signature as
$|a\ket_{S}$ ($a\in\{0, 1\}$).
A state that Alice sends at $t=0$ in Figure~\ref{Eve-Strategy1}
is given by
\[
\alpha L_{1}|0\ket_{Q} L_{2}|a\ket_{S}
+
\beta L_{1}|1\ket_{Q} L_{2}|a\oplus 1\ket_{S}.
\]
Bob measures only the system $S$ at $t=\tau$.
If he gets $|a\ket_{S}$,
Alice and Bob consider the state is transmitted correctly.
Bob uses the following projection operator for the measurement,
\[
\Pi^{QS}_{L_{1},L_{2}}
=
(L_{1}|0\ket\bra 0|L_{1}^{\dagger})_{Q}
\otimes
(L_{2}|a\ket\bra a|L_{2}^{\dagger})_{S}
+
(L_{1}|1\ket\bra 1|L_{1}^{\dagger})_{Q}
\otimes
(L_{2}|a\oplus 1\ket\bra a\oplus 1|L_{2}^{\dagger})_{S}.
\]

Here, we assume Eve makes an attack only on the qubit of the system $S$
as Figure~\ref{Eve-Strategy1}
($U$ and $V$ are arbitrary unitary transformations applied to one qubit).
We write the dynamical process of $S$
as a completely positive linear map
$\$$ that represents
Eve's Intercept/Resend attack on $S$\cite{Schumacher}.
Hence,
the density operator $\rho^{S}$ at $t=0$
evolves to $\$(\rho^{S})$ at $t=\tau$.
We can write the state of $QS$ at $t=\tau$ as
\beqa
\rho^{QS}_{L_{1},L_{2}}(\tau)
&=&
|\alpha|^{2}(L_{1}|0\ket\bra 0|L_{1}^{\dagger})_{Q}
\otimes{\$}(L_{2}|a\ket\bra a|L_{2}^{\dagger})_{S}\non \\
&&
+
\alpha\beta^{*}
(L_{1}|0\ket\bra 1|L_{1}^{\dagger})_{Q}
\otimes
{\$}(L_{2}|a\ket\bra a\oplus 1|L_{2}^{\dagger})_{S}\non \\
&&
+
\beta\alpha^{*}
(L_{1}|1\ket\bra 0|L_{1}^{\dagger})_{Q}
\otimes
{\$}(L_{2}|a\oplus 1\ket\bra a|L_{2}^{\dagger})_{S}\non \\
&&
+
|\beta|^{2}
(L_{1}|1\ket\bra 1|L_{1}^{\dagger})_{Q}
\otimes
{\$}(L_{2}|a\oplus 1\ket\bra a\oplus 1|L_{2}^{\dagger})_{S}.\non
\eeqa

We can write the probability $P$ that Bob obtains $|a\ket_{S}$ as
\beqa
P
&=&
\mbox{Tr}_{QS}
[\rho^{QS}_{L_{1},L_{2}}(\tau)\Pi^{QS}_{L_{1},L_{2}}]\non \\
&=&
|\alpha|^{2}
\bra a|L_{2}^{\dagger}
{\$}(L_{2}|a\ket\bra a|L_{2}^{\dagger})
L_{2}|a\ket
+
|\beta|^{2}
\bra a\oplus 1|L_{2}^{\dagger}
{\$}(L_{2}|a\oplus 1\ket\bra a\oplus 1|L_{2}^{\dagger})
L_{2}|a\oplus 1\ket.
\lab{Prob-Bob-1qubit-eve}
\eeqa
Seeing this,
we find the following fact.
Although the initial state $|\psi\ket_{Q}$ of the system $Q$
is a superposition of $|0\ket$ and $|1\ket$,
we may regard the state as a mixed state of $|\psi\ket_{Q}=|0\ket$
and $|\psi\ket_{Q}=|1\ket$
with classical probability for evaluating $P$.

\begin{figure}
\caption{Eve's Intercept/Resend attack on one qubit.}
\begin{center}
\includegraphics[scale=0.9]{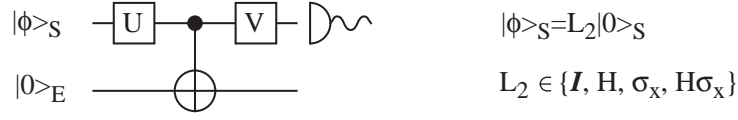}
\end{center}
\lab{eve-gatenet1}
\end{figure}

Therefore,
the probability of Bob's authentication is equal to
an average of probabilities
that
a network of quantum gates in Figure~\ref{eve-gatenet1}
gives $|\phi\ket_{S}$ as an outcome
from an incoming state $|\phi\ket_{S}=L_{2}|0\ket_{S}$
for all of $L_{2}\in{\cal L}$.
Here,
we evaluate $P$ as follows.
$U$ and $V$ are arbitrary unitary transformations applied to one qubit.
We assume
$U$ is defined as
\[
U|\varphi_{0}\ket=|0\ket,
\quad\quad
U|\varphi_{1}\ket=|1\ket,
\]
where $\{|\varphi_{0}\ket,|\varphi_{1}\ket\}$ is
a certain orthonormal basis of ${\cal H}_{2}$.
Then,
we write
$|\phi\ket=
c_{0}|\varphi_{0}\ket
+
c_{1}|\varphi_{1}\ket
$.
The state is transformed on the network of Figure~\ref{eve-gatenet1} as
\beqa
\lefteqn{
|\phi\ket_{S}|0\ket_{E}
=
(c_{0}|\varphi_{0}\ket_{S}+c_{1}|\varphi_{1}\ket_{S})|0\ket_{E}
} \non \\
&
\mapright{U}
&
(c_{0}|0\ket_{S}+c_{1}|1\ket_{S})|0\ket_{E} \non \\
&
\mapright{\mbox{\scriptsize C-NOT}}
&
c_{0}|0\ket_{S}|0\ket_{E}+c_{1}|1\ket_{S}|1\ket_{E} \non \\
&
\mapright{V}
&
c_{0} V|0\ket_{S}|0\ket_{E}+c_{1} V|1\ket_{S}|1\ket_{E}.
\lab{Eve-operation-eq1}
\eeqa
Seeing Eq.~(\ref{Eve-operation-eq1}),
we find that Eve measures the enciphered qubit in the basis
$\{|\varphi_{0}\ket,|\varphi_{1}\ket\}$
and sends a ket vector of a basis
$\{V|0\ket,V|1\ket\}$
according to a result of measurement.
Hence, we can write the probability $P_{\phi}$
that Bob gets the correct signature for $|\phi\ket_{S}$
in spite of Eve's illegal act as
\beq P_{\phi}=
|c_{0}|^{2}|\bra\phi|V|0\ket|^{2}
+
|c_{1}|^{2}|\bra\phi|V|1\ket|^{2}.
\lab{Prob-phi-1qbt-Eve}
\eeq

From now on,
for simplicity,
we write equations with density operators.
Defining
\[
\rho_{\phi}=|\phi\ket\bra\phi|,
\quad
\tilde{\rho}_{0}=|\varphi_{0}\ket\bra\varphi_{0}|,
\quad
\tilde{\rho}_{1}=|\varphi_{1}\ket\bra\varphi_{1}|,
\quad
\tilde{\rho}_{0}'=V|0\ket\bra 0|V^{\dagger},
\quad
\tilde{\rho}_{1}'=V|1\ket\bra 1|V^{\dagger},
\]
we can write Eq.~(\ref{Eve-operation-eq1}) as
\[
\rho_{\phi}
\rightarrow
{\$}(\rho_{\phi})
=
(\mbox{Tr}\rho_{\phi}\tilde{\rho}_{0})
\tilde{\rho}_{0}'
+
(\mbox{Tr}\rho_{\phi}\tilde{\rho}_{1})
\tilde{\rho}_{1}',
\]
and Eq.~(\ref{Prob-phi-1qbt-Eve}) as
\[
P_{\phi}=
\mbox{Tr}[{\$}(\rho_{\phi})\rho_{\phi}].
\]

Four density operators
$L_{2}|0\ket\bra 0|L_{2}^{\dagger}$
($L_{2}\in{\cal L}$),
emitted as $|\phi\ket_{S}$
with equal probability, are described as
\[
\rho_{\updownarrow}=\frac{1}{2}(\mbox{\boldmath $I$}+\sigma_{z}),
\quad
\rho_{\leftrightarrow}=\frac{1}{2}(\mbox{\boldmath $I$}-\sigma_{z}),
\quad
\rho_{\smallleftcircular}=\frac{1}{2}(\mbox{\boldmath $I$}+\sigma_{x}),
\quad
\rho_{\smallrightcircular}=\frac{1}{2}(\mbox{\boldmath $I$}-\sigma_{x}).
\]
Then
we define
\[
\tilde{\rho}_{i}
=\frac{1}{2}
[\mbox{\boldmath $I$}+(-1)^{i}
\mbox{\boldmath $X$}\cdot\mbox{\boldmath $\sigma$}],
\quad
\tilde{\rho}_{j}'
=\frac{1}{2}
[\mbox{\boldmath $I$}+(-1)^{j}
\mbox{\boldmath $X$}'\cdot\mbox{\boldmath $\sigma$}]
\quad
\mbox{for $i,j\in\{0,1\}$},
\]
where
$\mbox{\boldmath $X$}=(X,Y,Z)$
and
$\mbox{\boldmath $X$}'=(X',Y',Z')$
are arbitrary three-component real vectors
with
$|\mbox{\boldmath $X$}|^{2}=
|\mbox{\boldmath $X$}'|^{2}=1$.
Using the following formula,
\[
\mbox{Tr}
(\mbox{\boldmath $I$}+
\mbox{\boldmath $A$}\cdot\mbox{\boldmath $\sigma$})
(\mbox{\boldmath $I$}+
\mbox{\boldmath $B$}\cdot\mbox{\boldmath $\sigma$})
=
2(1+\mbox{\boldmath $A$}\cdot\mbox{\boldmath $B$}),
\]
and averaging four kinds of $P_{\phi}$,
we estimate $P_{B}$ that Bob measures the correct signature
in spite of Eve's illegal act at
\beq
P_{B}
=
\frac{1}{4}(P_{\updownarrow}+P_{\leftrightarrow}
+P_{\smallleftcircular}+P_{\smallrightcircular})
=
\frac{1}{4}(2+XX'+ZZ')
\leq\frac{3}{4}.
\eeq

Therefore,
the probability that Alice and Bob do not notice Eve
make an attack by the network of Figure~\ref{eve-gatenet1}
is $3/4$ or less.
If Eve makes the Intercept/Resend attack
on the system $Q$,
we can give a similar discussion.
Consequently,
the probability that Alice and Bob do not notice Eve's attacks
on $m$ qubits (either $Q$ or $S$ in each pair)
is $(3/4)^{m}$ at most.

\begin{figure}
\caption{Eve's Intercept/Resend attack on the system $Q$ and $S$.}
\begin{center}
\includegraphics[scale=0.9]{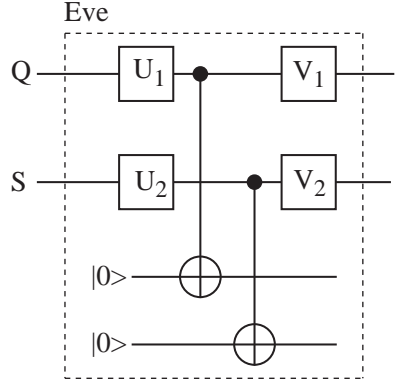}
\end{center}
\lab{Eve-Strategy2}
\end{figure}

Next,
we consider the case that
Eve makes the Intercept/Resend attack
on both qubits of a pair $QS$
independently in Figure~\ref{Eve-Strategy2}
($U_{1}$, $U_{2}$, $V_{1}$, and $V_{2}$ are
arbitrary unitary transformations).
Measuring $\rho^{QS}$ transmitted from Alice,
Eve sends the following density operator ${\rho^{QS}}'$ to Bob,
\[
{\rho^{QS}}'={\$}(\rho^{QS})
=\sum_{i,j\in\{0,1\}}
\mbox{Tr}
(\rho^{QS}
\tilde{\rho}_{Q,i}
\tilde{\rho}_{S,j}
)
\tilde{\rho}_{Q,i}'
\tilde{\rho}_{S,j}',
\]
where
\beq
\begin{array}{lclcl}
\tilde{\rho}_{Q,i}
&=&
U_{1}^{\dagger}|i\ket\bra i|U_{1}
&=&
(1/2)[\mbox{\boldmath $I$}+(-1)^{i}
\mbox{\boldmath $X$}_{1}\cdot\mbox{\boldmath $\sigma$}],\\
\tilde{\rho}_{S,j}
&=&
U_{2}^{\dagger}|j\ket\bra j|U_{2}
&=&
(1/2)[\mbox{\boldmath $I$}+(-1)^{j}
\mbox{\boldmath $X$}_{2}\cdot\mbox{\boldmath $\sigma$}],\\
\tilde{\rho}_{Q,i}'
&=&
V_{1}|i\ket\bra i|V_{1}^{\dagger}
&=&
(1/2)[\mbox{\boldmath $I$}+(-1)^{i}
\mbox{\boldmath $X$}_{3}\cdot\mbox{\boldmath $\sigma$}],\\
\tilde{\rho}_{S,j}'
&=&
V_{2}|j\ket\bra j|V_{2}^{\dagger}
&=&
(1/2)[\mbox{\boldmath $I$}+(-1)^{j}
\mbox{\boldmath $X$}_{4}\cdot\mbox{\boldmath $\sigma$}],\\
\end{array}
\eeq
and
$|\mbox{\boldmath $X$}_{k}|^{2}=1$
($k=1,\cdots,4$).

Here,
we can assume $|a\ket_{S}$ to be $|0\ket_{S}$ without losing generality.
We write a transmitted state of $Q$ as
$|\psi\ket_{Q}=\alpha|0\ket+\beta|1\ket$.
Because Alice has sixteen kinds of ways to send the state
for $L_{1}, L_{2}\in{\cal L}$,
Bob's final probability for authentication is described as
\beq
P_{B}=\frac{1}{16}\sum_{L_{1}, L_{2}\in{\cal L}}
\mbox{Tr}[\$(\rho^{QS}_{L_{1},L_{2}})\Pi^{QS}_{L_{1},L_{2}}].
\lab{Prob-Bob1}
\eeq
Writing
the density operator of the state
$\alpha L_{1}|0\ket_{Q}L_{2}|0\ket_{S}
+\beta L_{1}|1\ket_{Q}L_{2}|1\ket_{S}$
that Alice sends to Bob
as $\rho^{QS}_{L_{1},L_{2}}$,
we can describe its explicit form as
\beqa
\rho^{QS}_{L_{1},L_{2}}
&=&
|\alpha|^{2}
(L_{1}|0\ket\bra 0|L_{1}^{\dagger})_{Q}
\otimes
(L_{2}|0\ket\bra 0|L_{2}^{\dagger})_{S} \non \\
&&
+
\alpha\beta^{*}
(L_{1}|0\ket\bra 1|L_{1}^{\dagger})_{Q}
\otimes
(L_{2}|0\ket\bra 1|L_{2}^{\dagger})_{S}
+
\beta\alpha^{*}
(L_{1}|1\ket\bra 0|L_{1}^{\dagger})_{Q}
\otimes
(L_{2}|1\ket\bra 0|L_{2}^{\dagger})_{S} \non \\
&&
+
|\beta|^{2}
(L_{1}|1\ket\bra 1|L_{1}^{\dagger})_{Q}
\otimes
(L_{2}|1\ket\bra 1|L_{2}^{\dagger})_{S}.
\lab{Alice-message1}
\eeqa
The projection operator for Bob is given by
\[
\Pi^{QS}_{L_{1},L_{2}}
=
(L_{1}|0\ket\bra 0|L_{1}^{\dagger})_{Q}
\otimes
(L_{2}|0\ket\bra 0|L_{2}^{\dagger})_{S}
+
(L_{1}|1\ket\bra 1|L_{1}^{\dagger})_{Q}
\otimes
(L_{2}|1\ket\bra 1|L_{2}^{\dagger})_{S}.
\]
Eq.~(\ref{Prob-Bob1}) is linear for $\rho^{QS}_{L_{1},L_{2}}$.
Therefore,
we can divide Eq.~(\ref{Alice-message1}) into terms for calculation.

First,
we think about the first and forth terms
of Eq.~(\ref{Alice-message1}).
We write the first term as
\beq
\varrho^{QS}_{L_{1},L_{2}}
=
(L_{1}|0\ket\bra 0|L_{1}^{\dagger})_{Q}
\otimes
(L_{2}|0\ket\bra 0|L_{2}^{\dagger})_{S}.
\lab{Rho-diagonal-term}
\eeq
For example,
if $L_{1}=L_{2}=\mbox{\boldmath $I$}$,
we obtain
\beqa
\varrho^{QS}_{
\mbox{\scriptsize \boldmath $I$},
\mbox{\scriptsize \boldmath $I$}
}
&=&
(1/4)
(\mbox{\boldmath $I$}+\sigma_{z})_{Q}
\otimes
(\mbox{\boldmath $I$}+\sigma_{z})_{S}, \non \\
\Pi^{QS}_{
\mbox{\scriptsize \boldmath $I$},
\mbox{\scriptsize \boldmath $I$}
}
&=&
(1/4)
[
(\mbox{\boldmath $I$}+\sigma_{z})_{Q}
\otimes
(\mbox{\boldmath $I$}+\sigma_{z})_{S}
+
(\mbox{\boldmath $I$}-\sigma_{z})_{Q}
\otimes
(\mbox{\boldmath $I$}-\sigma_{z})_{S}
], \non
\eeqa
and
\[
\mbox{Tr}
[{\$}
(\varrho^{QS}_{
\mbox{\scriptsize \boldmath $I$},
\mbox{\scriptsize \boldmath $I$}
})
\Pi^{QS}_{
\mbox{\scriptsize \boldmath $I$},
\mbox{\scriptsize \boldmath $I$}
}
]
=(1/2)(1+Z_{1}Z_{2}Z_{3}Z_{4}).
\]
From similar calculations,
we get
\[
\mbox{Tr}
[{\$}
(\varrho^{QS}_{L_{1},L_{2}})
\Pi^{QS}_{L_{1},L_{2}}]
=
\left\{
\begin{array}{ll}
(1/2)(1+Z_{1}Z_{2}Z_{3}Z_{4})
&
\mbox{for $L_{1},L_{2}\in\{\mbox{\boldmath $I$},\sigma_{x}\}$},
\\
(1/2)(1+X_{1}X_{2}X_{3}X_{4})
&
\mbox{for $L_{1},L_{2}\in\{H,H\sigma_{x}\}$},
\\
(1/2)(1+Z_{1}X_{2}Z_{3}X_{4})
&
\mbox{for $L_{1}\in\{\mbox{\boldmath $I$},\sigma_{x}\}$,
$L_{2}\in\{H,H\sigma_{x}\}$},
\\
(1/2)(1+X_{1}Z_{2}X_{3}Z_{4})
&
\mbox{for $L_{1}\in\{H,H\sigma_{x}\}$,
$L_{2}\in\{\mbox{\boldmath $I$},\sigma_{x}\}$}.
\\
\end{array}
\right.
\]
Therefore,
we obtain
\[
\frac{1}{16}
\sum_{L_{1},L_{2}}
\mbox{Tr}
[{\$}
(\varrho^{QS}_{L_{1},L_{2}})
\Pi^{QS}_{L_{1},L_{2}}]
=
\frac{1}{2}
+\frac{1}{8}
(X_{1}X_{3}+Z_{1}Z_{3})
(X_{2}X_{4}+Z_{2}Z_{4}).
\]

Next,
we think about the second and third terms of Eq.~(\ref{Alice-message1}).
We write the second term as
\[
\Delta\varrho^{QS}_{L_{1},L_{2}}
=
(L_{1}|0\ket\bra 1|L_{1}^{\dagger})_{Q}
\otimes
(L_{2}|0\ket\bra 1|L_{2}^{\dagger})_{S}.
\]
For example,
if $L_{1}=L_{2}=\mbox{\boldmath $I$}$,
we obtain
\beqa
\Delta\varrho^{QS}_{
\mbox{\scriptsize \boldmath $I$},
\mbox{\scriptsize \boldmath $I$}
}
&=&
(1/4)
(\sigma_{x}+i\sigma_{y})_{Q}
\otimes
(\sigma_{x}+i\sigma_{y})_{S}, \non \\
\mbox{Tr}
[{\$}
(\Delta\varrho^{QS}_{
\mbox{\scriptsize \boldmath $I$},
\mbox{\scriptsize \boldmath $I$}
})
\Pi^{QS}_{
\mbox{\scriptsize \boldmath $I$},
\mbox{\scriptsize \boldmath $I$}
}
]
&=&(1/2)(X_{1}+iY_{1})(X_{2}+iY_{2})Z_{3}Z_{4}. \non
\eeqa
From similar calculations,
we get
\beqa
\lefteqn{
\mbox{Tr}
[{\$}
(\Delta\varrho^{QS}_{L_{1},L_{2}})
\Pi^{QS}_{L_{1},L_{2}}]
}\non \\
&&
\begin{array}{ccc}
& \epsilon=1 & \epsilon=-1 \\
=
\left\{
\begin{array}{r}
(1/2)[X_{1}+i\epsilon Y_{1}][X_{2}+i\epsilon Y_{2}]Z_{3}Z_{4}
\\
-(1/2)[X_{1}+i\epsilon Y_{1}][X_{2}-i\epsilon Y_{2}]Z_{3}Z_{4}
\\
(1/2)[Z_{1}-i\epsilon Y_{1}][Z_{2}-i\epsilon Y_{2}]X_{3}X_{4}
\\
-(1/2)[Z_{1}-i\epsilon Y_{1}][Z_{2}+i\epsilon Y_{2}]X_{3}X_{4}
\\
(1/2)[X_{1}+i\epsilon Y_{1}][Z_{2}-i\epsilon Y_{2}]Z_{3}X_{4}
\\
-(1/2)[X_{1}+i\epsilon Y_{1}][Z_{2}+i\epsilon Y_{2}]Z_{3}X_{4}
\\
(1/2)[Z_{1}-i\epsilon Y_{1}][X_{2}+i\epsilon Y_{2}]X_{3}Z_{4}
\\
-(1/2)[Z_{1}-i\epsilon Y_{1}][X_{2}-i\epsilon Y_{2}]X_{3}Z_{4}
\\
\end{array}
\right.
&
\begin{array}{c}
(L_{1},L_{2})=(\mbox{\boldmath $I$},\mbox{\boldmath $I$}),\\
(\mbox{\boldmath $I$},\sigma_{x}),\\
(H,H),\\
(H,H\sigma_{x}),\\
(\mbox{\boldmath $I$},H),\\
(\mbox{\boldmath $I$},H\sigma_{x}),\\
(H,\mbox{\boldmath $I$}),\\
(H,\sigma_{x}),\\
\end{array}
&
\begin{array}{c}
(\sigma_{x},\sigma_{x}),\\
(\sigma_{x},\mbox{\boldmath $I$}),\\
(H\sigma_{x},H\sigma_{x}),\\
(H\sigma_{x},H),\\
(\sigma_{x},H\sigma_{x}),\\
(\sigma_{x},H),\\
(H\sigma_{x},\sigma_{x}),\\
(H\sigma_{x},\mbox{\boldmath $I$}).\\
\end{array}
\\
\end{array}
\non
\eeqa
Consequently,
we obtain
\[
\frac{1}{16}
\sum_{L_{1},L_{2}}
\mbox{Tr}
[{\$}
(\Delta\varrho^{QS}_{L_{1},L_{2}})
\Pi^{QS}_{L_{1},L_{2}}]
=
-\frac{1}{8}Y_{1}Y_{2}
(Z_{3}-X_{3})
(Z_{4}-X_{4}).
\]

Finally, obtaining
\beq
P_{B}=
\frac{1}{2}
+
\frac{1}{8}
(X_{1}X_{3}+Z_{1}Z_{3})
(X_{2}X_{4}+Z_{2}Z_{4})
-
\frac{1}{8}
(\alpha\beta^{*}+\alpha^{*}\beta)
Y_{1}Y_{2}
(Z_{3}-X_{3})
(Z_{4}-X_{4}),
\lab{Prob-Bob-2qubit-Eve-attack}
\eeq
we can show $P_{B}\leq 3/4$.
(See Appendix~A.)
Therefore,
if $|\Psi\ket_{Q}$ is an $n$-qubit product state
and
if Eve makes the Intercept/Resend attack
on both qubits of a pair $QS$ independently
as Figure~\ref{Eve-Strategy2},
the probability that
Eve's illegal acts cannot be found
is equal to $3/4$ or less per one qubit.

Especially,
if the transmitted information is classical,
$|\Psi\ket_{Q}$
is a product state of $|0\ket$ and $|1\ket$.
All of the $2n$ qubits transmitted
are in states chosen from
four ket vectors of two conjugate bases
at random.
If we regard
$\mbox{\boldmath $a$}$ as a key of an $n$-bit random string
and $|\Psi\ket_{Q}$ as an $n$-bit enciphered classical message,
our method is equivalent to the one-time pad method with BB84.
Assuming Eve makes attacks on $m$ pairs of qubits in $QS$,
we can estimate the probability of Eve's success in eavesdropping
at $(3/4)^{m}$ or less.

Then,
we consider the case that
$|\Psi\ket_{Q}$ is an arbitrary entangled state of $n$ qubits.
The enciphered state of $|\Psi\ket_{Q}$
with $U_{\mbox{\scriptsize \boldmath $i$}}^{Q}$
is also entangled
and it is given by Eq.~(\ref{n-qbt-genrl-entg-state}).

First,
we consider that
Eve makes the Intercept/Resend attack
on either one
in a pair of qubits of the system $Q$ and $S$
as Figure~\ref{Eve-Strategy1}.
If Eve makes this attack on $m$ pairs
out of $n$ pairs of the system $QS$,
we can regard the transmission as
sending an ensemble of product states,
where each qubit is $|0\ket$ or $|1\ket$,
with classical probabilities,
like Eq.~(\ref{Prob-Bob-1qubit-eve}).
We can think in a similar way before
and conclude that
the probability Eve's illegal act cannot be found is $(3/4)^{m}$ or less.

Next,
we consider the case
that
Eve makes the attack on both qubits of a pair
on the entangled system $QS$
as Figure~\ref{Eve-Strategy2}.
If Eve makes this attack on $m$ pairs out of $n$ pairs,
we can write the probability Eve is not found
as an equation
which is similar to Eq.~(\ref{Prob-Bob-2qubit-Eve-attack})
and it is estimated at $(3/4)^{m}$ or less.
(See Appendix~B.)

\section{Privacy amplification process}\lab{Privacy-Amplification}

From previous discussion, we obtain the following results.
If an arbitrary quantum state is enciphered by our method,
a probability that Alice and Bob do not notice Eve is
at most $3/4$ per one qubit.
(Both product and entangled states are available.
We assume that Eve always makes eavesdropping
with the Intercept/Resend attack.)
Hence, if Eve makes attacks on $m$ enciphered pairs,
her success probability is given by $(3/4)^{m}$
and it decreases exponentially against $m$.

However, there is a problem.
In our method,
if Eve replaces a pair of enciphered qubits
with a pair of random ones,
Alice and Bob do not notice her illegal act with probability of $1/2$.
In this case,
they disclose passwords and Eve obtains
one qubit of original information with fidelity $1$.
It is important that Eve gets a correct qubit
and she knows that she obtains the correct one.

Such a problem can be also occurred in the BB84.
It is possible that Alice and Bob share the same random binary string
and Eve knows a few bits of it exactly.
To overcome this trouble, for example,
Alice and Bob can choose some bits at random
from the shared binary string
and make a new bit from a summation of them
with modulo $2$\cite{bennett-BB84exp}.
If they repeat this process
and create a new binary string that is shorter than original one,
Eve's expected information decreases to $0$ in some asymptotic limit.
Such a technique is called privacy amplification.

On the other hand, in our protocol,
Eve's success probability for eavesdropping on one-qubit
cannot always reach $0$.
To decrease it to $0$ asymptotically,
Alice and Bob apply our protocol over and over again.
To make discussion simple,
we consider encryption of one-qubit quantum information for a while.

Preparing an arbitrary one-qubit state
$|\psi_{1}\ket$
and a one-qubit signature
$|a_{1}\ket$ ($\forall a_{1} \in\{0,1\}$),
Alice applies our protocol to
$|\psi_{1}\ket |a_{1}\ket$
and generates an entangled two-qubit state
$|\psi_{2}\ket$.
Then, she prepares other qubits
$|a_{2}\ket |a_{3}\ket$
for a signature,
and enciphers
$|\psi_{2}\ket |a_{2}\ket |a_{3}\ket$
again.
She obtains a four-qubit state
$|\psi_{3}\ket$.

If Eve wants to get quantum information of
$|\psi_{1}\ket$
with fidelity $1$,
she has to interact with all four qubits of
$|\psi_{3}\ket$.
For example,
if Eve replaces
$|\psi_{3}\ket$
with random four qubits,
Alice and Bob notice her illegal act with probability of
$(1/2)^{3}=1/8$,
because they carry out the authentication process with
$|a_{1}\ket$, $|a_{2}\ket$, and
$|a_{3}\ket$.

If Alice enciphers
$|\psi_{1}\ket$
for $n$ times,
the $n$th encryption needs $2^{n-1}$ signature qubits.
If Eve makes attacks on all enciphered qubits of
$|\psi_{n+1}\ket$,
the probability that Alice and Bob do not notice her act is
$(3/4)^{N/2}$ at most,
where $N=2^{n}$ is the number of all enciphered qubits.
(The probability $3/4$ comes from the fact
that Alice enciphers the state
with rectilinear and circular bases at random,
and it does not depend on
$|\psi_{n}\ket$.)
Hence, Eve's success probability
that she gets
$|\psi_{1}\ket$ with fidelity of $1$
decreases exponentially against the number of qubits,
and reaches $0$ in the limit of
$N \rightarrow \infty$.

Another method is as follows.
Alice and Bob share a random binary string beforehand
as the first password (subscripts of Pauli matrices)
in secrecy  by BB84.
Because they do not need to disclose it,
Eve can never get information of $|\psi_{1}\ket$ at all
even in the case that they do not notice Eve's disturbance.
In this method,
the privacy amplification has to be done for BB84 actually.

\section{Discussion}\lab{CONCLUSION}

To understand our method more clearly,
we consider a simple one and compare it with ours.
For transmitting an $n$-qubit quantum state in secrecy,
we can take the following method.
Enciphering an $n$-qubit state
$\rho_{n}$ as Eq.~(\ref{n-qubit-cry-mix}),
Alice prepares other $n$ qubits as check ones that are given as
$\{|0\ket,|1\ket\}$
or
$\{(1/\sqrt{2})(|0\ket\pm|1\ket)\}$
at random respectively.
Then, Alice permutes all of the $2n$ qubits at random
and sends them to Bob.

Here, we assume
Eve tries to eavesdrop on only one qubit of $\rho_{n}$.
Because Eve does not know which qubits are check ones,
a probability that Alice and Bob fail to notice Eve's illegal act
can be
$(1/2)[1+(3/4)]=7/8$
as the maximum,
in spite of $3/4$ for our method.
This is because we use entanglement in our method.

Even if Eve prepares an arbitrary one-qubit state by herself
in spite of taking away a qubit from $\rho_{n}$,
its expectation value of fidelity is equal to $1/2$.
This shows that Eve's success probability of eavesdropping is
always equal to $1/2$ or more.
In our method, if Eve interacts enciphered qubits,
a probability of her success is equal to at most $3/4$
per one qubit (without privacy amplification).
It is similar to the BB84.

\begin{figure}
\caption{Eve's attack on the system $QS$ by using entanglement.}
\begin{center}
\includegraphics[scale=0.9]{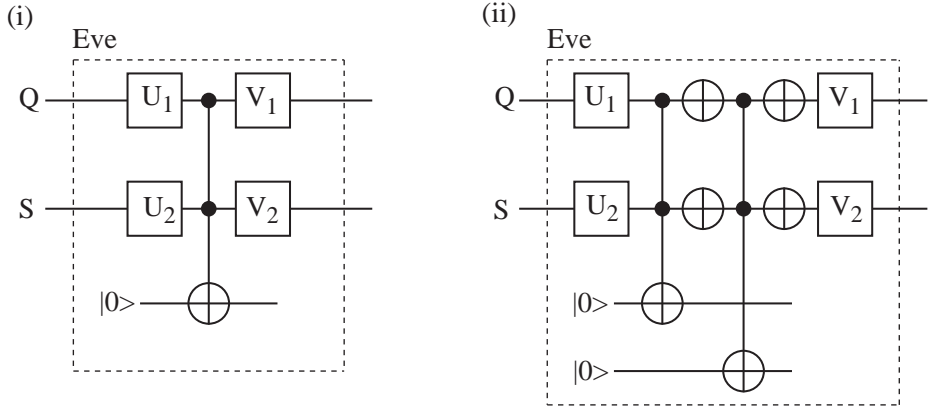}
\end{center}
\lab{Eve-Strategy3}
\end{figure}

In this paper,
the security against Eve's attack
with using entanglement
is not considered
(for example, a case that she uses a quantum computer
for eavesdropping as Figure~\ref{Eve-Strategy3}).
In Figure~\ref{Eve-Strategy3},
it is difficult to evaluate the upper bound of the probability that
Bob measures the signature correctly
for arbitrary unitary transformations
$U_{1}$, $U_{2}$, $V_{1}$, and $V_{2}$.
For instance,
assuming
\[
U_{1}=U_{2}=V_{1}=V_{2}=U
\quad\quad
\mbox{where }
U=
\left(
\begin{array}{cc}
\cos(\pi/8) & \sin(\pi/8) \\
\sin(\pi/8) & -\cos(\pi/8) \\
\end{array}
\right),
\]
and
$|\Psi\ket_{Q}$ represents classical information
(an $n$-qubit product state of $|0\ket$ and $|1\ket$),
we get
$P_{B}=(13/16)>(3/4)$ for Figure~\ref{Eve-Strategy3}.~i
and
$P_{B}=(11/16)<(3/4)$ for Figure~\ref{Eve-Strategy3}.~ii.
Eavesdropping with $U$
is equivalent to
measuring and resending a qubit in the following basis,
\[
|\varphi_{i}\ket\bra\varphi_{i}|=
\frac{1}{2}
(
\mbox{\boldmath $I$}
+(-1)^{i}
\mbox{\boldmath $X$}\cdot
\mbox{\boldmath $\sigma$}
),
\quad\quad
\mbox{\boldmath $X$}
=
(\frac{1}{\sqrt{2}},0,\frac{1}{\sqrt{2}}),
\]
which is called the Breidbart basis\cite{bennett-BB84exp}.
About Figure~\ref{Eve-Strategy3}.~i,
$P_{B}$ may exceed $3/4$.
But,
the amount of information Eve can extract
in Figure~\ref{Eve-Strategy3}.~i
seems to be
less than the amount of information
she
obtains by the Intercept/Resend attack
on one qubit as Figure~\ref{Eve-Strategy1}.

In our method,
if Eve takes away qubits,
Alice and Bob lose original information of them.

We mentioned in Sec.~\ref{introduction}
that
the classical key distribution can be done only by the uncertainty,
and the entanglement is not essential
for it\cite{Ekert91}\cite{BBM92_VSEkert91}.
On the other hand,
for transmitting quantum states by the quantum teleportation,
the entanglement plays
an essential role\cite{QTeleportOrg}.
Our method uses both two properties.

Recently,
the method for transmitting classical binary data
(not a classical random string)
in secrecy
with a pair of entangled photons
has been proposed\cite{Shimizu-Imoto}.
It is characterized by the following facts.
First,
Alice and Bob prepare two conjugate bases on ${\cal H}_{2}^{2}$ each
for encoding message and measuring photons.
Second,
they use a two-dimensional subspace of ${\cal H}_{2}^{2}$
for encoding a binary digit.

\bigskip
\noindent
{\bf \large Acknowledgements}
\smallskip

We thank O.~Hirota, M.~Osaki, and H.~Inamori
for helpful discussions.
We also thank R.~de~Wolf for useful comments.
H.~A. thanks M.~Okuda for encouragement.

\appendix
\section{The maximum value of $P_{B}$ for a product state $|\Psi\ket_{Q}$}

Here, we show that
$P_{B}$ defined in Eq.~(\ref{Prob-Bob-2qubit-Eve-attack})
never exceeds $3/4$.

Because of $|\alpha|^{2}+|\beta|^{2}=1$,
we get $-1\leq\alpha\beta^{*}+\alpha^{*}\beta\leq1$.
Hence,
altering the signs of $X_{i}$
as $X_{i}\rightarrow-X_{i}$ ($i=1,\cdots,4$),
we can write the upper bound of $P_{B}$ as
\[
P_{B}\leq\frac{1}{2}+\frac{1}{8}f_{MAX},
\]
where $f_{MAX}$ is the maximum value of
\beq
f=
(X_{1}X_{3}+Z_{1}Z_{3})(X_{2}X_{4}+Z_{2}Z_{4})
+Y_{1}Y_{2}(X_{3}+Z_{3})(X_{4}+Z_{4}),
\lab{F-function-XYZ}
\eeq
with
$|\mbox{\boldmath $X$}_{i}|^{2}=1$
and
$X_{i},Y_{i},Z_{i}\geq 0$ for $i=1,\cdots,4$.

Seeing Eq.~(\ref{F-function-XYZ}),
we give another form of $f$ as follows,
\[
f=|\mbox{\boldmath $A$}\cdot\mbox{\boldmath $B$}|,
\]
where
\beqa
\mbox{\boldmath $A$}
&=&
(X_{1}X_{3}+Z_{1}Z_{3}, Y_{1}(X_{3}+Z_{3})), \non \\
\mbox{\boldmath $B$}
&=&
(X_{2}X_{4}+Z_{2}Z_{4}, Y_{2}(X_{4}+Z_{4})). \non
\eeqa
(We pay an attention that
$\mbox{\boldmath $A$}$ and $\mbox{\boldmath $B$}$
are two-component real vectors.)
From the Cauchy-Schwarz inequality,
we get
\[
f\leq |\mbox{\boldmath $A$}||\mbox{\boldmath $B$}|.
\]
Therefore, by estimating the maximum values of
$|\mbox{\boldmath $A$}|$ and $|\mbox{\boldmath $B$}|$,
we derive the upper bound of $f$.

We can write $|\mbox{\boldmath $A$}|^{2}$
in the following form,
\[
|\mbox{\boldmath $A$}|^{2}
=
(1-Z_{1}^{2})X_{3}^{2}
+(1-X_{1}^{2})Z_{3}^{2}
+2(X_{1}Z_{3})(Z_{1}X_{3})
+2Y_{1}^{2}(X_{3}Z_{3}).
\]
On the other hand,
from the arithmetic-geometric inequality,
we obtain
\beqa
(X_{1}Z_{3})(Z_{1}X_{3})
&\leq&
\frac{1}{2}[(X_{1}Z_{3})^{2}+(Z_{1}X_{3})^{2}], \non \\
X_{3}Z_{3}
&\leq&
\frac{1}{2}(X_{3}^{2}+Z_{3}^{2}). \non
\eeqa
Therefore, we get
\beq
|\mbox{\boldmath $A$}|^{2}
\leq
(1+Y_{1}^{2})(X_{3}^{2}+Z_{3}^{2})
\leq
1+Y_{1}^{2}
\leq
2.
\eeq
We obtain $|\mbox{\boldmath $A$}|\leq \sqrt{2}$.
In a similar way,
we obtain $|\mbox{\boldmath $B$}|\leq \sqrt{2}$.
From these results,
we can conclude that $f\leq 2$
and $f_{MAX}=2$.

\section{The maximum value of $P_{B}$ for an entangled state $|\Psi\ket_{Q}$}

We estimate the probability
that Eve's illegal act cannot be found
in the case where she makes the Intercept/Resend attack
on $m$ pairs of qubits on the system $QS$
for an arbitrary entangled $|\Psi\ket_{Q}$ of $n$ qubits.

For simplicity,
we assume $|\Psi\ket_{Q}$ to be an arbitrary entangled state
of two-qubit system $qq'$ at first,
\[
|\Psi\ket_{Q}
=\sum_{i,j\in\{0,1\}}
c_{ij}
|i\ket_{q}
\otimes
|j\ket_{q'}\in\forall{\cal H}_{2}^{2}.
\]
Alice puts two qubits of the system $S$($=ss'$)
for the signature
on
the qubits of the system $Q$($=qq'$) respectively.
Then,
she makes entanglement between the systems $Q$ and $S$
with C-NOT gates,
applies $L_{1}$, $L_{2}$, $L_{1}'$, $L_{2}'\in{\cal L}$
to four qubits $q$, $s$, $q'$, $s'$ respectively,
and sends them to Bob (see Figure~\ref{Eve-Strategy1}).
Eve makes the Intercept/Resend attacks
on the systems $q$, $s$, $q'$, $s'$ respectively
as shown in Figure~\ref{Eve-Strategy2}.
We can assume the initial states of qubits $s$, $s'$ that represents the signature
to be $|0\ket_{s}|0\ket_{s'}$ without losing generality.

Writing the state sent by Alice
as
\[
\sum_{i,j\in\{0,1\}}c_{ij}
L_{1}|i\ket_{q}L_{2}|i\ket_{s}
\otimes
L_{1}'|j\ket_{q'}L_{2}'|j\ket_{s'},
\]
we can describe the density operator
explicitly as
\beqa
\rho^{QS}_{L_{1}L_{2}L_{1}'L_{2}'}
&=&
(L_{1}L_{2}L_{1}'L_{2}')
\sum_{i,j\in\{0,1\}}
[
|c_{ij}|^{2}
(|i\ket\bra i|_{q}\otimes|i\ket\bra i|_{s})
\otimes
(|j\ket\bra j|_{q'}\otimes|j\ket\bra j|_{s'}) \non \\
&&
\quad
+
c_{ij}c^{*}_{i\bar{j}}
(|i\ket\bra i|_{q}\otimes|i\ket\bra i|_{s})
\otimes
(|j\ket\bra \bar{j}|_{q'}\otimes|j\ket\bra \bar{j}|_{s'}) \non \\
&&
\quad
+
c_{ij}c^{*}_{\bar{i}j}
(|i\ket\bra \bar{i}|_{q}\otimes|i\ket\bra \bar{i}|_{s})
\otimes
(|j\ket\bra j|_{q'}\otimes|j\ket\bra j|_{s'}) \non \\
&&
\quad
+
c_{ij}c^{*}_{\bar{i}\bar{j}}
(|i\ket\bra \bar{i}|_{q}\otimes|i\ket\bra \bar{i}|_{s})
\otimes
(|j\ket\bra \bar{j}|_{q'}\otimes|j\ket\bra \bar{j}|_{s'})
]
(L_{1}L_{2}L_{1}'L_{2}')^{\dagger},
\eeqa
where
$\bar{i}=i+1\mbox{ (mod $2$)}$.

Eavesdropping on the state
$\rho^{QS}_{L_{1}L_{2}L_{1}'L_{2}'}$,
Eve transforms it to the following state,
\[
{\$}(\rho^{QS}_{L_{1}L_{2}L_{1}'L_{2}'})
=\sum_{i,j,k,l\in\{0,1\}}
\mbox{Tr}
(\rho^{QS}_{L_{1}L_{2}L_{1}'L_{2}'}
\tilde{\rho}_{q,i}\tilde{\rho}_{s,j}
\tilde{\rho}_{q',k}\tilde{\rho}_{s',l})
\tilde{\rho}_{q,i}'\tilde{\rho}_{s,j}'
\tilde{\rho}_{q',k}'\tilde{\rho}_{s',l}'.
\]
where
\beq
\begin{array}{lllllll}
\tilde{\rho}_{q,i}
&=&
(1/2)[\mbox{\boldmath $I$}+(-1)^{i}
\mbox{\boldmath $X$}_{1}\cdot\mbox{\boldmath $\sigma$}],
&\quad&
\tilde{\rho}_{s,j}
&=&
(1/2)[\mbox{\boldmath $I$}+(-1)^{j}
\mbox{\boldmath $X$}_{2}\cdot\mbox{\boldmath $\sigma$}],\\
\tilde{\rho}_{q,i}'
&=&
(1/2)[\mbox{\boldmath $I$}+(-1)^{i}
\mbox{\boldmath $X$}_{3}\cdot\mbox{\boldmath $\sigma$}],
&\quad&
\tilde{\rho}_{s,j}'
&=&
(1/2)[\mbox{\boldmath $I$}+(-1)^{j}
\mbox{\boldmath $X$}_{4}\cdot\mbox{\boldmath $\sigma$}],\\
\tilde{\rho}_{q',k}
&=&
(1/2)[\mbox{\boldmath $I$}+(-1)^{k}
\mbox{\boldmath $X$}'_{1}\cdot\mbox{\boldmath $\sigma$}],
&\quad&
\tilde{\rho}_{s',l}
&=&
(1/2)[\mbox{\boldmath $I$}+(-1)^{l}
\mbox{\boldmath $X$}'_{2}\cdot\mbox{\boldmath $\sigma$}],\\
\tilde{\rho}_{q',k}'
&=&
(1/2)[\mbox{\boldmath $I$}+(-1)^{k}
\mbox{\boldmath $X$}'_{3}\cdot\mbox{\boldmath $\sigma$}],
&\quad&
\tilde{\rho}_{s',l}'
&=&
(1/2)[\mbox{\boldmath $I$}+(-1)^{l}
\mbox{\boldmath $X$}'_{4}\cdot\mbox{\boldmath $\sigma$}],\\
\end{array}
\eeq
and
$|\mbox{\boldmath $X$}_{k}|^{2}=
|\mbox{\boldmath $X$}_{k}'|^{2}=1$
($k=1,\cdots,4$).
Bob measures it with the projection operator,
\beqa
\Pi^{QS}_{L_{1}L_{2}L_{1}'L_{2}'}
&=&
(L_{1}L_{2})
(|0\ket\bra 0|_{q}\otimes|0\ket\bra 0|_{s}
+
|1\ket\bra 1|_{q}\otimes|1\ket\bra 1|_{s}
)(L_{1}L_{2})^{\dagger}
\non \\
&&
\quad
\otimes
(L_{1}'L_{2}')(|0\ket\bra 0|_{q'}\otimes|0\ket\bra 0|_{s'}
+
|1\ket\bra 1|_{q'}\otimes|1\ket\bra 1|_{s'}
)(L_{1}'L_{2}')^{\dagger}.\non
\eeqa

The probability that
Bob measures the correct signature
is given by
\beqa
P_{B}
&=&
(\frac{1}{16})^{2}
\sum_{L_{1},L_{2}\in{\cal L}}
\:
\sum_{L_{1}',L_{2}'\in{\cal L}}
\mbox{Tr}
[{\$}(\rho^{QS}_{L_{1}L_{2}L_{1}'L_{2}'})
\Pi^{QS}_{L_{1}L_{2}L_{1}'L_{2}'}]
\non \\
&=&
[\frac{1}{2}+\frac{1}{8}(X_{1}X_{3}+Z_{1}Z_{3})(X_{2}X_{4}+Z_{2}Z_{4})]
[\frac{1}{2}+\frac{1}{8}(X'_{1}X'_{3}+Z'_{1}Z'_{3})(X'_{2}X'_{4}+Z'_{2}Z'_{4})]
\non \\
&&
+\sum_{i,j\in\{0,1\}}
\{
c_{ij}c^{*}_{i\bar{j}}
[\frac{1}{2}+\frac{1}{8}(X_{1}X_{3}+Z_{1}Z_{3})(X_{2}X_{4}+Z_{2}Z_{4})]
\non \\
&&\quad\quad\quad\quad\quad\quad
\times
[-\frac{1}{8}Y'_{1}Y'_{2}(Z'_{3}-X'_{3})(Z'_{4}-X'_{4})]
\non \\
&&
+
c_{ij}c^{*}_{\bar{i}j}
[-\frac{1}{8}Y_{1}Y_{2}(Z_{3}-X_{3})(Z_{4}-X_{4})]
[\frac{1}{2}+\frac{1}{8}(X'_{1}X'_{3}+Z'_{1}Z'_{3})(X'_{2}X'_{4}+Z'_{2}Z'_{4})]
\non \\
&&
+
c_{ij}c^{*}_{\bar{i}\bar{j}}
[-\frac{1}{8}Y_{1}Y_{2}(Z_{3}-X_{3})(Z_{4}-X_{4})]
[-\frac{1}{8}Y'_{1}Y'_{2}(Z'_{3}-X'_{3})(Z'_{4}-X'_{4})]
\}.
\non
\eeqa
From
$\sum_{i,j\in\{0,1\}}|c_{ij}|^{2}=1$,
we obtain
$|\sum_{i,j\in\{0,1\}}c_{ij}c^{*}_{i\bar{j}}|\leq 1$,
$|\sum_{i,j\in\{0,1\}}c_{ij}c^{*}_{\bar{i}j}|\leq 1$,
and
\\
$|\sum_{i,j\in\{0,1\}}c_{ij}c^{*}_{\bar{i}\bar{j}}|\leq 1$.
Therefore,
using the result obtained in Appendix~A,
we can conclude
\[
P_{B}\leq
(\frac{1}{2}+\frac{1}{8}f_{MAX})^{2}
=(\frac{3}{4})^{2}.
\]
(We pay attention to a fact
that each term of $P_{B}$ can be gathered with a binomial coefficient.)
When Eve attacks on $m$ pairs
out of enciphered qubits generated
from an arbitrary $n$-qubit entangled state $|\Psi\ket_{Q}$,
we obtain
$P_{B}\leq(3/4)^{m}$.

\end{document}